# Improved Mean-Field Scheme for the Hubbard Model

J. C. Hicks, J. Tinka Gammel,*

*Materials Research Branch, Code 573 NCCOSC/RDT&E Division, San Diego, California 92152-5000*

(August 23, 1994)

## Abstract

Ground state energies and on-site density-density correlations are calculated for the 1-D Hubbard model using a linear combination of the Hubbard projection operators. The mean-field coefficients in the resulting linearized Equations of Motion (EOM) depend on both one-particle static expectation values as well as static two-particle correlations. To test the model, the one particle expectation values are determined self-consistently while using Lanczos determined values for the two particle correlation terms. Ground state energies and on-site density-density correlations are then compared as a function of $U$ to the corresponding Lanczos values on a 12 site Hubbard chain for 1/2 and 5/12 fillings. To further demonstrate the validity of the technique, the static correlation functions are also calculated using a similar EOM approach, which ignores the effective vertex corrections for this problem, and compares those results as well for a 1/2 filled chain. These results show marked improvement over standard mean-field techniques.

1993 PACS: 71.10.+x, 71.27.+a

Typeset using REVTEX

Following the discovery of high $T_c$ superconductivity in the copper oxides [1], strongly correlated systems have generated a great deal of interest in condensed-matter theory. It is recognized that the electronic properties of many systems, *e.g.* the heavy-Fermion systems and high $T_c$ superconductors, are basically determined by strong correlations [2]. The fundamental model describing a tight binding system with strong local correlations is the Hubbard model [3], characterized by the hopping matrix element, $t$, and the on-site interaction energy, $U$. Despite the simple nature of the model (basically a one parameter model, $U/t$), it is in general not analytically tractable. Interestingly enough the model has been solved exactly for the 1/2-filled band in one dimension [4]. Solutions have also been obtained under certain limiting conditions. In the limit of large $U$, allowing for only single occupancy at each site, expansions in $1/N$, where $N$ is the spin degeneracy have been solved [5] as well as an exact mean-field solution in the limit of large dimensions [6]. However, in spite of the physical insight gained from these approaches, these 'exact' solutions are clearly approximations to the actual physical model which involves a spin degeneracy of 2 and dimensionality ranging from 1 to 3. Thus, a full understanding of the physics described by the Hubbard model has so far escaped theoretical physicists. What we wish to describe in this communication is a straightforward mean-field scheme that linearizes the Equations of Motion (EOM) that result from the Hubbard model and then compare the resulting fits to exact numerical Lanczos solutions for a 12 site system. We start with the charge conjugation symmetric version of the Hubbard model,

$$H = -\sum_{ij\sigma} t_{ij} c_{i\sigma}^\dagger c_{j\sigma} + \frac{U}{2} \sum_{i\sigma} (n_{i\sigma} - \frac{1}{2})(n_{i-\sigma} - \frac{1}{2}). \tag{1}$$

The EOM obtained from this model using finite temperature formalism are well known and are given by

$$(-\frac{\partial}{\partial \tau} + \mu + \frac{U}{2}) c_{i\sigma} + \sum_j t_{ij} c_{i\sigma} - U n_{i-\sigma} c_{i\sigma} = 0 \tag{2a}$$

$$(-\frac{\partial}{\partial \tau} + \mu - \frac{U}{2}) n_{i-\sigma} c_{i\sigma} = -\sum_j t_{ij} n_{i-\sigma} c_{j\sigma}$$
$$- \sum_j t_{ij} (c_{i-\sigma}^\dagger c_{j-\sigma} - c_{j-\sigma}^\dagger c_{i-\sigma}) c_{i\sigma}. \tag{2b}$$

The standard mean-field approximations (which were also employed by Hubbard [3]) are $n_{i-\sigma} c_{j\sigma} \approx \langle n_{i-\sigma} \rangle c_{j\sigma}$, and $(c_{i-\sigma}^\dagger c_{j-\sigma} - c_{j-\sigma}^\dagger c_{i-\sigma}) c_{i\sigma} \approx \langle c_{i-\sigma}^\dagger c_{j-\sigma} - c_{j-\sigma}^\dagger c_{i-\sigma} \rangle c_{i\sigma}$. Due to translational invariance of the system, the second or off-site term is ignored and the mean field value of $\langle n_{i-\sigma} \rangle$ is given by $n_e/2$ where $n_e$ represents the electronic filling of the system. The resulting mean-field solution demonstrates the short range Coulomb-generated energy gap and the corresponding splitting into two bands but is well known to exhibit several shortcomings, the most obvious being the absence of a Mott transition for all dimensions.

We have decided to expand the usual mean-field linearization by projecting out $n_{i-\sigma} c_{i\sigma}$ terms as well as terms proportional to $c_{i\sigma}$. To do this, we form a linearly independent combination of the Hubbard projection operators first described by one of the authors [7]. Our projection operators then take on the form,

$$\hat{P}\hat{O} = \sum_{i\sigma} \frac{\langle\{(1-n_{i-\sigma})c^\dagger_{i\sigma}, \hat{O}\}\rangle}{\langle 1-n_{i-\sigma}\rangle} c_{i\sigma} + \sum_{i\sigma} \frac{\langle\{(n_{i-\sigma}-\langle n_{i-\sigma}\rangle)c^\dagger_{i\sigma}, \hat{O}\}\rangle}{\langle 1-n_{i-\sigma}\rangle\langle n_{i-\sigma}\rangle} n_{i-\sigma}c_{i\sigma}. \tag{3}$$

This definition of the projection operator satisfies $\hat{P}c_{i\sigma} = c_{i\sigma}$ and $\hat{P}n_{i-\sigma}c_{i\sigma} = n_{i-\sigma}c_{i\sigma}$, and thus is consistent with the idempotent property of projection operators, $\hat{P}^2 = \hat{P}$. Additionally, it incorporates terms in $c_{i\sigma}$ and $n_{i-\sigma}c_{i\sigma}$ that the original mean-field scheme ignored. Equation 2a above is unchanged but the new mean-field approximation for 2b now reads,

$$\sum_j t_{ij}\langle n_{i-\sigma}\rangle c_{j\sigma} + (-\frac{\partial}{\partial\tau} + \mu - \frac{U}{2})n_{i-\sigma}c_{i\sigma} \approx$$
$$+\delta(\langle n_{i-\sigma}\rangle c_{i\sigma} - n_{i-\sigma}c_{i\sigma}) + \sum_j \delta t_{ij}(\langle n_{j-\sigma}\rangle c_{j\sigma} - n_{j-\sigma}c_{j\sigma}) \tag{4a}$$

where, for a translationally invariant system,

$$\delta = \sum_j t_{ij} \frac{\langle n_{i\sigma}(c^\dagger_{i-\sigma}c_{j-\sigma} + c^\dagger_{j-\sigma}c_{i-\sigma})\rangle - \langle c^\dagger_{j-\sigma}c_{i-\sigma}\rangle}{\langle n_{i-\sigma}\rangle\langle 1-n_{i-\sigma}\rangle}, \tag{4b}$$

and

$$\delta t_{ij} = t_{ij} \frac{\langle \delta n_{i-\sigma}\delta n_{j-\sigma}\rangle - \langle c^\dagger_{j\sigma}c_{i\sigma}(c^\dagger_{i-\sigma}c_{j-\sigma} + c^\dagger_{j-\sigma}c_{i-\sigma})\rangle}{\langle n_{j-\sigma}\rangle\langle 1-n_{j-\sigma}\rangle}. \tag{4c}$$

Using the standard definitions for the Green's functions in this problem, $g_{ij\sigma}(\tau) = -\langle T_\tau c_{i\sigma}(\tau)c^\dagger_{j\sigma}(0)\rangle$, and $f_{ij\sigma}(\tau) = -\langle T_\tau n_{i-\sigma}(\tau)c_{i\sigma}(\tau)c^\dagger_{j\sigma}(0)\rangle$, the mean-field EOM for the Green's functions are conveniently expressed as

$$\sum_j \left[H^{MF}_{ij}\right] \begin{bmatrix} g_{ij\sigma}(\tau) \\ f_{ij\sigma}(\tau) \end{bmatrix} = \begin{bmatrix} 1 \\ \langle n_{i-\sigma}\rangle \end{bmatrix} \delta_{ij}\delta(\tau) \tag{5a}$$

where $H^{MF}_{ij}$ is given by

$$\left[H^{MF}_{ij}\right] = \begin{bmatrix} (-\frac{\partial}{\partial\tau} + \mu + \frac{U}{2})\delta_{ij} + t_{ij} & -U\delta_{ij} \\ \langle n_{i-\sigma}\rangle t_{ij} - \langle n_{j-\sigma}\rangle \delta t_{ij} - \delta\delta_{ij} & (-\frac{\partial}{\partial\tau} + \mu - \frac{U}{2})\delta_{ij} + \delta t_{ij} \end{bmatrix}. \tag{5b}$$

The coefficient for the on-site term, $\delta$, can be expressed as simple linear combination of a single particle expectation values for $f$ and its equivalent charge-conjugate expression for holes. However, the difficulty in obtaining mean-field solutions to the Hubbard model using this approach is in obtaining values for the static two particle correlations that occur as coefficients for the off-site terms in the linearization process. These correlation coefficients demonstrate the direct coupling that occurs between the mean-field solution and the multiparticle correlations that arise in strongly correlated systems. These correlation functions can be calculated using a standard linear response formalism by taking variations of the EOM with respect to a density field, yielding density-density correlations, or to a bond field, yielding bond-bond correlations. The resulting two-particle correlations are then coupled to three-particle correlations. Rather than extend this heirarchy, we have decided to demonstrate the significant improvement of this technique, as compared to the original mean-field approach, by using numerically obtained Lanczos wave functions to calculate the two-particle correlations. With the coefficients for the off-site terms fixed, this mean-field problem can then be solved self-consistently. The solutions for both $g$ and $f$ are then used to predict ground state energy, $E_g$, and the on-site density-density correlation, $\langle n_{i\sigma} n_{i-\sigma} \rangle$, for a 12 site chain. These values, as well as those obtained for the original Hubbard solution, are shown in Figs.1(a) and 1(b) as a function of $U/t$ for a 1/2 filled system and in Figs. 2(a) and 2(b) for a 5/12 filled system. The significant improvement of this mean-field scheme is especially noticeable for the 1/2 filled case. The difference in the ground state energy between the "Hubbard mean-field" (HMF) and and the Lanczos (L) solution is approximately 1/2 the difference between the standard mean-field (MF) and the Lanczos result (albeit with a different sign). The improvement in the on-site correlations is even more significant with the difference between the HMF solution and the exact result being approximately 1/3 the difference between the MF solution and the Lanczos result. Since the MF solutions themselves for 5/12 filling are significantly improved as compared to the 1/2 filled case, the observed improvements of the HMF solutions for this filling are not as impressive but are none-the-less significant, particularly for the on-site density correlations.

To further validate the improvement in this mean-field scheme, we can again use the EOM approach to calculate the static correlation functions that were used as coefficients in this mean-field approach and compare them as well to the Lanczos results. This approach ignores all vertex corrections and consequently the three particle correlations that we alluded to above. However, for the sake of simplicity, it can further demonstrate the effectiveness of this approach. First, we define two pairs of coupled time-dependent correlation functions;

$$G_{nmij\sigma}(\tau_1 \tau_2) = -\langle T_\tau (c^\dagger_{n-\sigma} c_{m-\sigma})_{\tau_1} c_{i\sigma}(\tau_2) c^\dagger_{j\sigma}(0) \rangle, \tag{6a}$$

and,

$$F_{nmij\sigma}(\tau_1 \tau_2) = -\langle T_\tau (c^\dagger_{n-\sigma} c_{m-\sigma})_{\tau_1} (n_{i-\sigma} c_{i\sigma})_{\tau_2} c^\dagger_{j\sigma}(0) \rangle, \tag{6b}$$

as well as;

$$\mathcal{G}_{nmij\sigma}(\tau_1 \tau_2) = -\langle T_\tau (c^\dagger_{n\sigma} c_{m\sigma})_{\tau_1} c_{i\sigma}(\tau_2) c^\dagger_{j\sigma}(0) \rangle, \tag{6c}$$

and,

$$\mathcal{F}_{nmij\sigma}(\tau_1 \tau_2) = -\langle T_\tau (c^\dagger_{n\sigma} c_{m\sigma})_{\tau_1} (n_{i-\sigma} c_{i\sigma})_{\tau_2} c^\dagger_{j\sigma}(0) \rangle. \tag{6d}$$

Taking $\tau$ derivatives with respect to $\tau_2$ of the above correlation functions leads to sets of EOM very similar to those for the Green's functions,

$$\sum_l \left[ H_{il}^{MF}(\tau_2) \right] \begin{bmatrix} G_{nmlj\sigma}(\tau_1 \tau_2) \\ F_{nmlj\sigma}(\tau_1 \tau_2) \end{bmatrix} = \begin{bmatrix} g_{nm\sigma}(0^-) \\ \mathcal{G}_{nmii\sigma}(\tau_1 0^-) \end{bmatrix} \delta_{ij} \delta(\tau_2) +$$

$$\begin{bmatrix} 0 \\ G_{niij\sigma}(\tau_1^+ \tau_1) \delta_{mi} - G_{imij\sigma}(\tau_1^+ \tau_1) \delta_{ni} \end{bmatrix} \delta(\tau_1 - \tau_2), \tag{7a}$$

and

$$\sum_l \left[ H_{il}^{MF}(\tau_2) \right] \begin{bmatrix} \mathcal{G}_{nmlj\sigma}(\tau_1\tau_2) \\ \mathcal{F}_{nmlj\sigma}(\tau_1\tau_2) \end{bmatrix} = \begin{bmatrix} g_{nm\sigma}(0^-) \\ G_{nmii\sigma}(\tau_1 0^-) \end{bmatrix} \delta_{ij}\delta(\tau_2) -$$

$$\begin{bmatrix} g_{mj\sigma}(\tau_1) \\ G_{iimj\sigma}(\tau_1^+\tau_1) \end{bmatrix} \delta_{ni}\delta(\tau_1-\tau_2). \tag{7b}$$

Inverting these EOM to obtain the time-dependent response or correlation functions yields a set of expressions from which we can evaluate the static correlation functions required for our Hubbard mean-field scheme. However, before we proceed it is interesting to note some of the self-consistencies that occur with these EOM, most notable among them include;

$$\mathcal{G}_{iiij\sigma}(\tau\tau^-) \equiv -\langle T_\tau (n_{i\sigma} c_{i\sigma})_\tau c_{j\sigma}^\dagger(0) \rangle = 0, \tag{8a}$$

$$\mathcal{G}_{iiij\sigma}(\tau\tau^+) \equiv -\langle T_\tau (c_{i\sigma} n_{i\sigma})_\tau c_{j\sigma}^\dagger(0) \rangle = g_{ij\sigma}(\tau), \tag{8b}$$

and

$$G_{iiii\sigma}(00^-) \equiv \langle n_{i-\sigma} n_{i\sigma} \rangle = f_{ii\sigma}(0^-). \tag{8c}$$

Thus diagonal components of the correlation functions reduce to the correct Green's functions in this EOM scheme. Meanwhile, the density-density correlation that we require can be written,

$$\langle \delta n_{i+1-\sigma} \delta n_{i-\sigma} \rangle = \mathcal{G}_{i+1i+1ii-\sigma}(00^-) - \langle n_{-\sigma} \rangle^2 =$$

$$- \left[ \frac{g_{ii+1-\sigma}(0^-) g_{i+1i-\sigma}(0^+) + \langle n_\sigma \rangle \langle 1-n_\sigma \rangle {}_{12}H^{-1}_{ii+1}(0^-) {}_{12}H^{-1}_{i+1i}(0^+)}{1 - {}_{12}H^{-1}_{ii}(0^-)^2} \right] \tag{9}$$

while the bond-bond correlation of interest reduces to,

$$\langle c_{i+1\sigma}^\dagger c_{i\sigma} (c_{i-\sigma}^\dagger c_{i+1-\sigma} + c_{i+1-\sigma}^\dagger c_{i-\sigma}) \rangle =$$

$$G_{i+1ii+1i\sigma}(00^-) + G_{i+1iii+1\sigma}(00^-) = -2\langle \delta n_{i+1-\sigma} \delta n_{i-\sigma} \rangle. \tag{10}$$

Here ${}_{12}H^{-1}_{ij}$ is the 12 component of the inverse of the mean-field Hamiltonian operator. These expressions for the static correlations of interest can easily be evaluated from the mean-field solutions. These results can also be compared to the correlations obtained from Lanczos wave functions. These results for a half-filled 12 site band are shown in Fig.3. They begin to show some differences, albeit with similar trends, for U on the order of the band width. This is probably largely due to the omission of relevant vertex corrections which become important at larger $U$'s, rather than being due to significant errors in the mean-field scheme as presented. In fact, self-consistent mean-field results can be obtained using this scheme with the above expressions being used for the correlations. The results are virtually identical to those presented here for $U \leq 4t$. The point we wish to make here is that this mean-field scheme for the Hubbard model represents a significant improvement over standard mean-field approximations and can be easily, as well as accurately, applied for $U$ less than or equal to the bandwidth ($U \leq 4t$). The authors also feel that by including vertex corrections while ignoring 3-particle fluctuations, this mean-field scheme can answer many questions for the nearly half-filled Hubbard model, including the Mott transition in a 3-dimensions. Additionally, this approach can easily be extended to include nearest-neighbor repulsion effects arising from $V_{ij}$.

The authors acknowledge Andrei Ruckenstein for suggesting this problem and the Office of Naval Research for supporting this work.

FIGURES

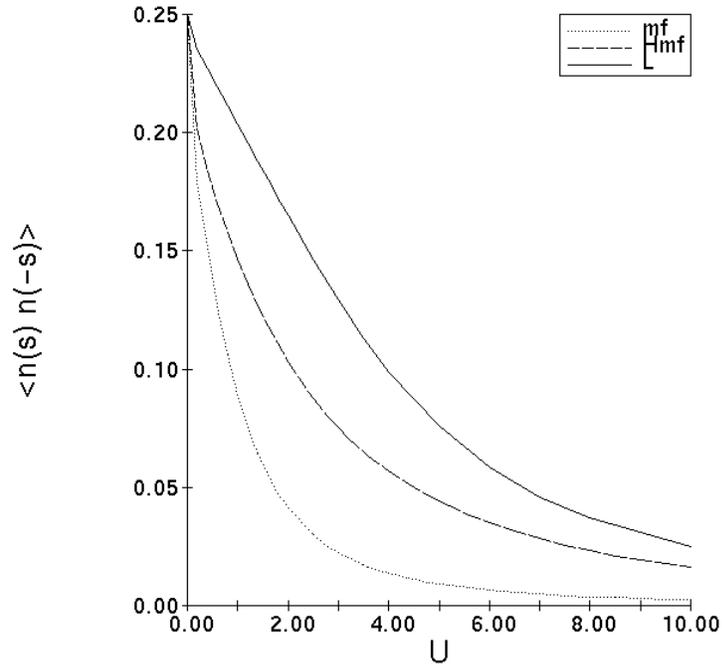

Figure 1a. On-site charge correlations as a function of U for a 12 site 1/2 filled band. $t=1$ is used in all figures.

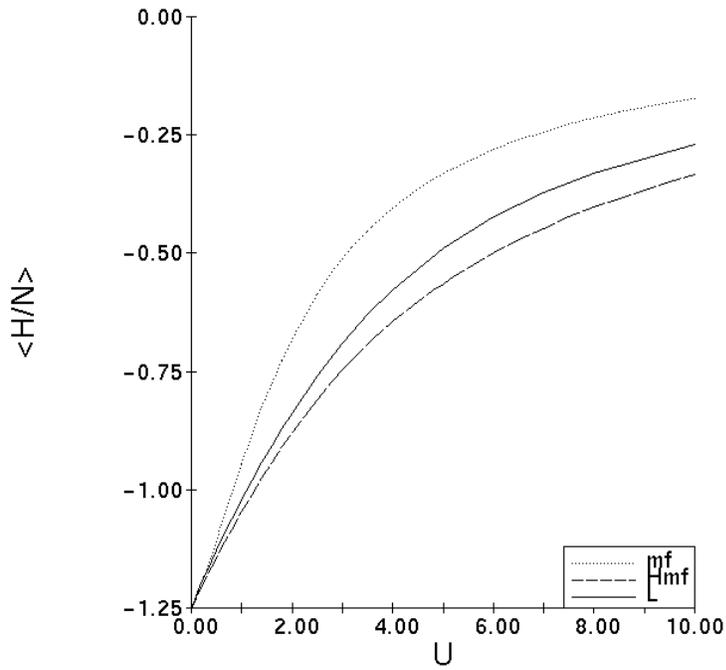

Figure 1b. Ground-state energy as a function of U for a 12 site 1/2 filled band.

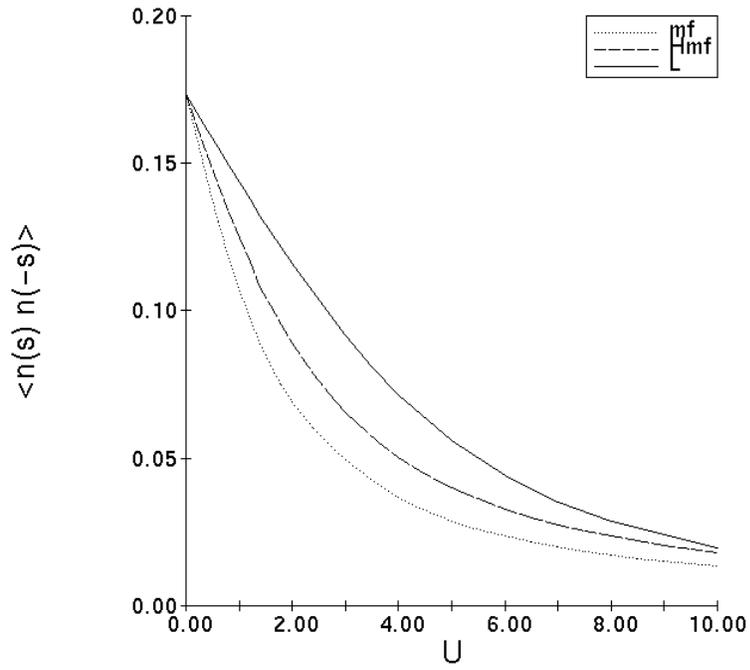

Figure 2a. On-site charge correlations as a function of U for a 12 site 5/12 filled band.

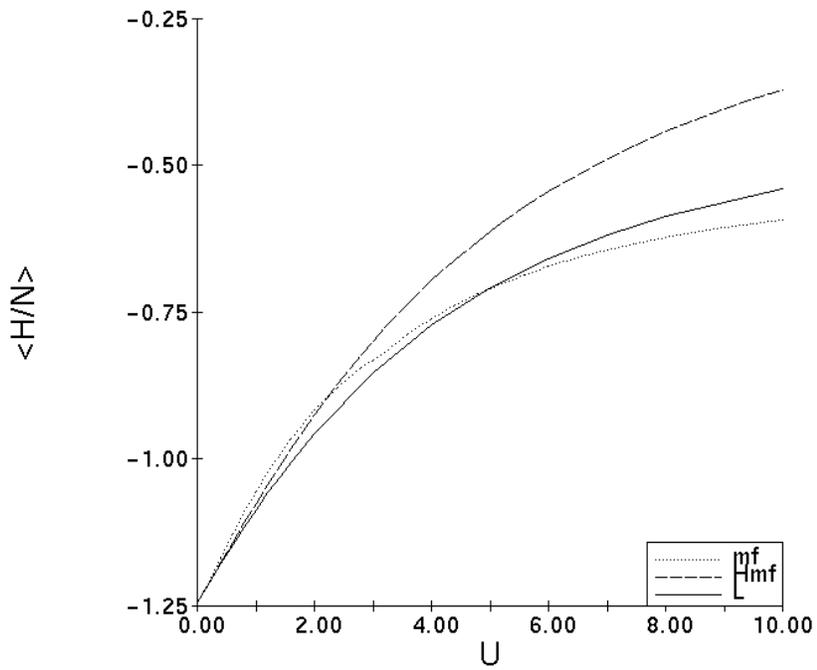

Figure 2b. Ground-state energy as a function of U for a 12 site 5/12 filled band.

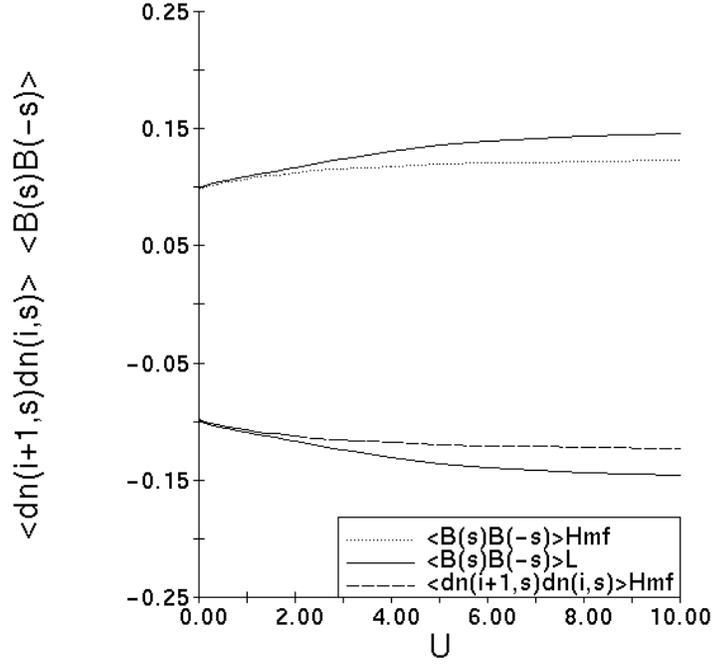

Figure 3. Density-Density correlations and Bond-Bond $(B(s) = (c^\dagger_{i+1s}c_{is} + c^\dagger_{is}c_{i+1s})/2)$ correlations as a function of U for a 1/2 filled band.

@10